\documentclass[11pt,a4paper]{article}
\usepackage[utf8]{inputenc}
\usepackage[english]{babel}
\usepackage{amsmath}
\usepackage{amsfonts}
\usepackage{amssymb}
\usepackage{graphicx}
\usepackage{bbm}
\usepackage{svg}
\usepackage{sidecap}
\usepackage{wrapfig}
\usepackage{algorithm}
\usepackage[noend]{algpseudocode}
\usepackage{color, colortbl}
\usepackage{tabularx}
\usepackage{array}	
\usepackage[left=2cm,right=2cm,top=2cm,bottom=2cm]{geometry}
\definecolor{Gray}{gray}{0.9}
\definecolor{LightCyan}{rgb}{0.88,1,1}
\newcolumntype{L}[1]{>{\raggedright\let\newline\\\arraybackslash\hspace{0pt}}m{#1}}
\newcolumntype{R}[1]{>{\raggedleft\let\newline\\\arraybackslash\hspace{0pt}}m{#1}}
\date{}
\title{From Correlation to Causation: Estimation of Effective Connectivity from Continuous Brain Signals based on Zero-Lag Covariance}
\author{Jonathan Schiefer, Alexander Niederbühl, Volker Pernice, Carolin Lennartz, Pierre LeVan, \\ Jürgen Henning and Stefan Rotter}

\begin{document}

\maketitle
\section*{Abstract}
Knowing brain connectivity is of great importance both in basic research and for clinical applications. 
We are proposing a method to infer directed connectivity from zero-lag covariances of neuronal activity recorded at multiple sites. This allows us to identify causal relations that are reflected in neuronal population activity.
To derive our strategy, we assume a generic linear model of interacting continuous variables, the components of which represent the activity of local neuronal populations. 
The suggested method for inferring connectivity from recorded signals exploits the fact that the covariance matrix derived from the observed activity contains information about the existence, the direction and the sign of connections. 
Assuming a sparsely coupled network, we disambiguate the underlying causal structure via $L^1$-minimization, which is known to prefer sparse solutions.
In general, this method is suited to infer effective connectivity from resting state data of various types. 
We show that our method is applicable over a broad range of structural parameters regarding network size and connection probability of the network. We also explored parameters affecting its activity dynamics, like the eigenvalue spectrum.
Also, based on the simulation of suitable Ornstein-Uhlenbeck processes to model BOLD dynamics, we show that with our method it is possible to estimate directed connectivity from zero-lag covariances derived from such signals.
In this study, we consider measurement noise and unobserved nodes as additional confounding factors. 
Furthermore, we investigate the amount of data required for a reliable estimate.
Additionally, we apply the proposed method on full-brain resting-state fast fMRI datasets. 
The resulting network exhibits a tendency for close-by areas being connected as well as inter-hemispheric connections between corresponding areas. 
In addition, we found that a surprisingly large fraction of more than one third of all identified connections were of inhibitory nature.

\section*{Introduction}
The networks of the brain are key to understanding its function and dysfunction \cite{Li2017}.
Depending on the methods employed to assess structure and to record activity, networks may be defined at different levels of resolution.
Their nodes may be individual neurons, linked by chemical or electrical synapses.
Alternatively, nodes may also be conceived as populations of neurons, with links represented by the net effect of all synaptic connections that exist between two populations.
In any case, this defines the structural substrate of brain connectivity, representing the physical (causal) basis of neuronal interactions.
Nodes in a brain network influence each other by sending signals.
For example, the activities of nodes in a network are generally not independent, and neuronal dynamics are characterized by correlations among the nodes involved in the network.
This suggests an alternative perspective on active brain networks: Functional connectivity assigns a link to a pair of nodes to the degree to which their activities are correlated.
It has been argued that this concept emphasizes connections that ``matter'', including the possibility that the same substrate may give rise to different networks, depending on how they are used. 
As a consequence, functional connectivity and structural connectivity are not equivalent.
A well-known phenomenon is that two nodes may be correlated, even if there is no direct anatomical link between them.
For example, a shared source of input to both nodes may generate such a correlation, which does not correspond to a direct interaction between the two nodes.
Apart from that, correlation is a symmetric relation between two nodes, whereas a physical connection implies a cause-effect relation that is directed.
There have, in fact, been multiple attempts to overcome the shortcomings of functional connectivity, especially the lack of directed interaction.
The term \textit{effective connectivity} has been suggested for this \cite{Friston2011}.
The idea is to bring the networks, inferred from activity measurements, closer to structural connectivity, which can only be inferred with anatomical methods.
The dichotomy between structural and functional aspects of connectivity raises the general question whether it is possible to infer brain networks from recorded activity.
We are only beginning to understand the forward link between structural connectivity and functional connectivity.
As a consequence, it is possible to compute correlations from connectivity in certain simplified network scenarios \cite{Pernice2011}.
The correspondence between connectivity and correlation, however, is not one-to-one.
Networks with different connectivity may lead to exactly the same correlations between nodes. As a consequence, the inverse problem of inferring connectivity from correlation is generally ill-defined.
As we will demonstrate in this paper, additional assumptions about the connectivity can help to resolve the ambiguity. 
Specifically, we search for the network with the lowest number of nonzero edges (via $L^1$-minimization) to disambiguate the problem. 
Structural, functional and effective connectivity are not equally well accessible.
Some aspects of the anatomical structure can be assessed \textit{post mortem} by invasive tracing methods, or non-invasively by Diffusion Tensor Imaging, DTI.
In contrast, functional connectivity is based on statistical relationships between the activity of neuronal populations and can be easily estimated from recorded signals.
For estimating effective connectivity there are methods like Dynamic Causal Modelling, DCM \cite{Havlicek2017, Stephan2010}, Granger causality \cite{Smith2010} and others \cite{Freestone2014, Gilson2016, Ting2015, Roebroeck2011, MP2017, Marrelec2006, Timme2014}.
Only few methods to infer effective connectivity, however, can deal with large numbers of nodes ($40$ or more) based on zero-lag correlation only.
However, they are either limited to small networks \cite{Gates2012}, or to directed acyclic graphs \cite{Ramsey2016}. 
Here, we are proposing a new method for the estimation of effective connectivity from population activity in the brain, especially BOLD-related signals.
The new method is a variant of the procedure described in \cite{Pernice2013}, based on a $L^1$-minimization. 
For the method proposed here it is sufficient to use zero-lag covariances to estimate directed effective connectivity.

\section*{Materials and methods}
\subsection*{Estimation method}\label{methods:model}
The main idea of our estimation method is inspired by the finding, ``that the key to determining the direction of the causal relationship between $X$ and $Y$ lies in `the presence of a third variable $Z$ that correlates with $Y$ but not with X,' as in the collider $X \rightarrow Y \leftarrow Z$ \dots'' \cite{Pearl2000, Rebane1987}. 

\begin{figure*}[!h]
\centering
\includegraphics[scale=0.45]{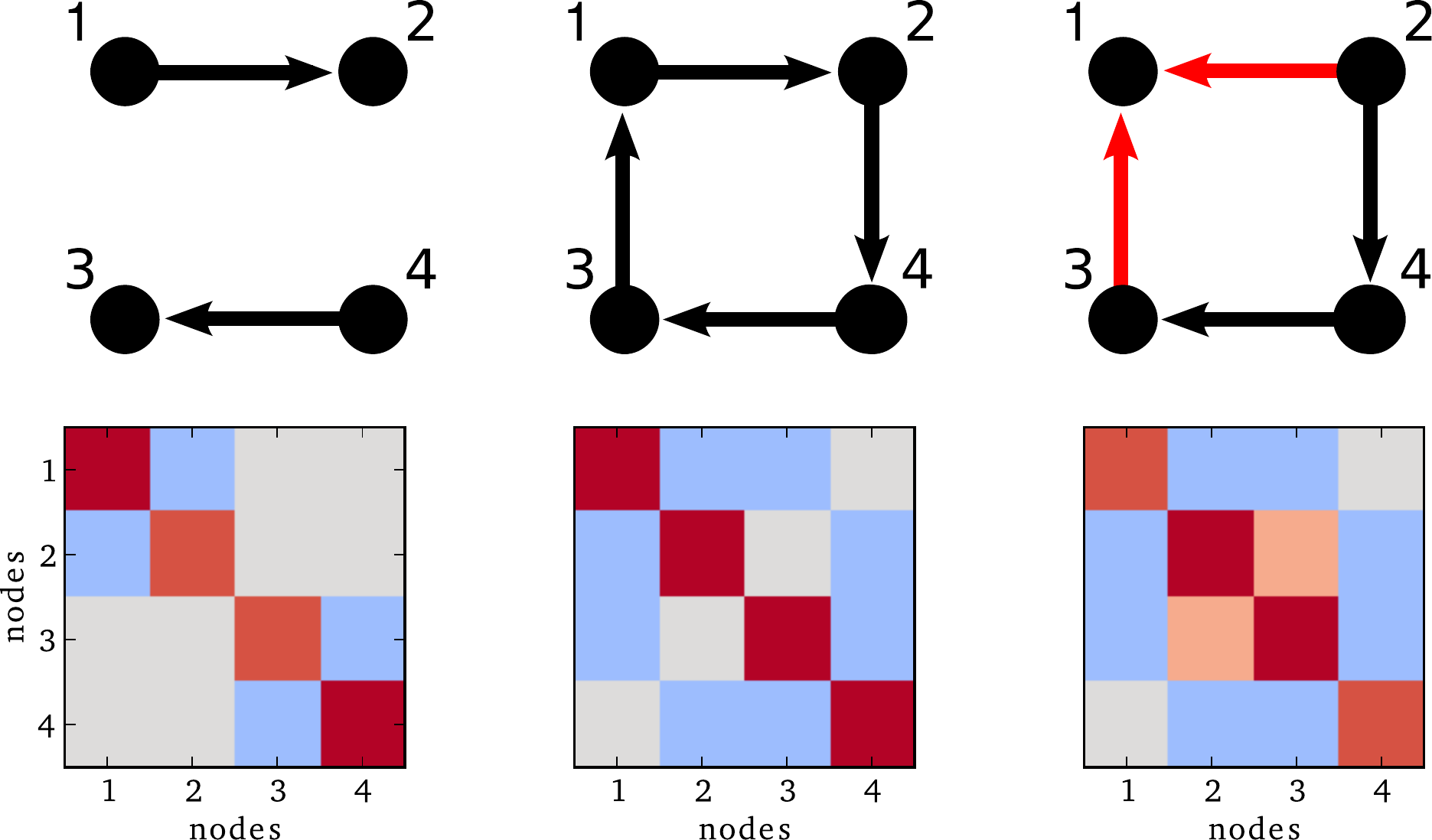}
\caption{ Collider structures are encoded in the inverse covariance matrix. Upper row: Three simple network architectures. Lower row: The corresponding inverse covariance matrices, red color represents positive entries, blue color stands for negative ones. In the left and middle column, the entries $(2,3)$ and $(3,2)$ are $0$. The only difference between the right column and the middle column is that the connection between node $1$ and $2$ is flipped, such that nodes $1$, $2$ and $3$ form a collider structure. Although there is still only an indirect connection between node $2$ and $3$, the entry in the corresponding inverse covariance matrix is now non-zero.}\label{fig:collider}
\end{figure*}

Similarly, assuming a linear interaction model, the presence of a collider structure in a network (see Fig~\ref{fig:collider}) produces specific entries in the corresponding inverse covariance (precision) matrix. 
Fig~\ref{fig:collider} shows a disconnected network in the left column, and a network which induces the same covariance matrix if all links have opposite direction in the middle column.
In the latter case an estimation of the direction is impossible, because there is simply no information about it in the covariance matrix.
Whenever a collider structure is present, however, the entry in the inverse covariance matrix for the two source nodes (here, $2$ and $3$) is non-zero. 
This is due to the fact that in a linear model the entry in the inverse covariance matrix depends not only on the connections of the nodes $2$ and $3$, but also whether these nodes have a common target.
This means the presence of a collider structure allows us to disambiguate the direction of this particular connection.

We consider here a scenario, where the interaction between nodes is described by a generic linear model. Assuming stationarity, let the neural activity $x(t)$ be implicitly defined by the consistency equation
\begin{equation}\label{eq:activity}
x (t ) = (G \ast x ) (t ) + v (t ) 
\end{equation} 
where $G(t)$ is a matrix of causal interaction kernels and $v(t)$ denotes fluctuating external inputs (``driving noise'').
All variables are also listed in table~\ref{table_variables}.
Fourier transformation of Eq~(\ref{eq:activity}) yields
\begin{equation*}
\hat{x} ( f ) = \hat{G} ( f )\hat{x} ( f ) + \hat{v} ( f )
\end{equation*}
and simple rearrangement leads to
\begin{equation*} \label{eq:activity_fourier}
\hat{x} ( f ) = \left( \mathbbm{1} - \hat{G} ( f ) \right)^{-1}  \hat{v} ( f )
\end{equation*}
where $\hat{x}$ denotes the Fourier transform of $x$.
The cross spectral density of the signals is then given by 
\begin{equation*}
\hat{C} ( f ) = \left( \mathbbm{1} - \hat{G} ( f ) \right)^{-1} \hat{Z} ( f ) \left( \mathbbm{1} - \hat{G}^* ( f ) \right)^{-1}
\end{equation*}
where $\hat{Z}(f)$ is the cross-spectral density of the external inputs. It follows 
\begin{align}\label{eq:cinv}
\hat{C}^{-1} ( f ) &= \left( \mathbbm{1} - \hat{G}^* ( f ) \right) \hat{Z}^{-1} ( f ) \left( \mathbbm{1} - \hat{G} ( f ) \right) \nonumber \\ &= B^* ( f) B ( f ) 
\end{align}
with $B ( f ) = \sqrt{\hat{Z}^{-1} ( f )} \left( \mathbbm{1} - \hat{G} ( f ) \right)$.
In our model, we assume that the components of the external fluctuating input are pairwise stochastically independent for all nodes. 
Then, $\hat{Z}$ is a diagonal matrix, and we make the additional assumption that $\hat{Z} = \mathbbm{1}$.
For the linear model considered here, there is a relation between covariance and connectivity, which can be exploited for the estimation of connectivity from correlation. 
In the case $\hat{Z} = \mathbbm{1}$ it is given by 
\[ \hat{C}^{-1}(f) = \left( \mathbbm{1} - \hat{G}^* ( f ) \right)\left( \mathbbm{1} - \hat{G} ( f ) \right) = \mathbbm{1} - \hat{G}(f) -  \hat{G}^* ( f ) + \hat{G}^* ( f ) \hat{G}( f ) 
\]
where the last term contributes the information of the collider structures. 
If the matrix product $\hat{G}^* ( f ) \hat{G}( f )$ has a non-zero off-diagonal entry the corresponding nodes have outgoing connections terminating at the same node, which means these nodes form a collider.

It is clear that for any unitary matrix $U \in \mathcal{U}(n)$ the product $UB$ is still a solution of Eq (\ref{eq:cinv}), as $U^*U = \mathbbm{1}$. 
We will resolve this ambiguity with an $L^{1}$ minimization which is known to prefer sparse solutions under certain conditions \cite{Candes2005}. 
In order to find $G$ from a given $C$ we first fix an initial matrix $B$, and then search for a unitary matrix $U \in \mathcal{U}(n)$ such that $\| UB \|_1$ is minimal, so we are minimizing the function 
\begin{align}\label{map:gamma}
\Gamma \colon &\mathcal{U}(n) \longrightarrow \mathbb{R} \nonumber \\
&U \longmapsto \| UB \|_1 .
\end{align}

\begin{table}[!ht]
\centering
\caption{
{\bf Variables used for the estimation method and simulation}}
\begin{tabularx}{\textwidth}{L{14.5cm} R{1.5cm} }
\multicolumn{1}{X}{\bf Variable~name}& \multicolumn{1}{>{\raggedleft}X}{\bf Symbol} \\ 
\rowcolor{Gray}
node activity & $x(t)$ \\ 
network connectivity & $G$ \\ 
\rowcolor{Gray}
external inputs & $v(t)$ \\ 
cross-spectral density & $\hat{C}(f)$ \\ 
\rowcolor{Gray}
covariance of external input & $\hat{Z}(f)$ \\ 
unitary matrix & $U$ \\ 
\rowcolor{Gray}
$L^1$-norm cost function & $\Gamma$ \\ 
gradient & $d$ \\
\rowcolor{Gray}
initial matrix & $B_0$ \\ 
Wiener process & $w(t)$\\ 
\rowcolor{Gray}
stationary covariance matrix & $\sigma$\\ 
simulation step & $\Delta t$\\ 
\rowcolor{Gray}
time constant of activity & $\tau$\\ 
regularisation-controlling parameter for regularized ICOV & $\lambda$
\end{tabularx}
\label{table_variables}
\end{table}

\subsubsection*{Gradient descent}\label{methods:gd}

To estimate the connectivity matrix from the covariance matrix we use a conjugate gradient descent algorithm similar to \cite{Abrudan2009, Wen2013} for minimizing the function $\Gamma(U)$ given in Eq~(\ref{map:gamma}), implemented in Python. 
For details please see supporting information, algorithm~\ref{algo:conj}.
For the gradient \[ d_{ij} =  \frac{\partial \Gamma (U)}{\partial U_{i,j}}  \] of the cost function $\Gamma(U)$, $a = \frac{1}{2}(d-d^\ast)$ is skew-hermitian, and the matrix exponential of a skew-hermitian matrix is unitary. 
This means, starting in a point $U_\mathrm{act}$ and choosing an appropriate step size $\delta$, we obtain a point $U_\mathrm{new} = \exp(- \delta a ) U_\mathrm{act}$ with $\Gamma( U_{\mathrm{new}}) < \Gamma( U_\mathrm{act}) $. 
In other words, the new point has a smaller $L^1$-norm than the old one and still satisfies the condition $\hat{C}^{-1} = B^* U_\mathrm{new}^* U_\mathrm{new} B$. 
Iterating this procedure until convergence leads to a point with locally minimal $L^1$-norm.  

The two conditions for convergence are inspired by \cite{Wen2013}. The first one is a condition on the norm of the gradient. In each step, it is checked if
\[
\| d - U d^* U \|_F = \sqrt{\sum_{i,j} |d_{ij} (U d^* U)_{ij} |^2} < \mathrm{gtol}
\]
is fulfilled, where $\| \ldots \|_F$ is the Frobenius norm and $\mathrm{gtol} > 0$ is the convergence tolerance.
As a second (alternative) condition, it is checked whether simultaneously
\[
\frac{\| U - U_\mathrm{old} \|_F}{\sqrt{N}} < \mathrm{xtol}
\qquad \text{and} \qquad 
\frac{|\Gamma(U_\mathrm{old}) - \Gamma(U)|}{|\Gamma(U_\mathrm{old} + 1)|} < \mathrm{ftol}
\]
are fulfilled.
The values used are listed in table \ref{table_parameter_est}.
Before convergence the cost function typically oscillates around a certain value. To avoid stopping at a random phase of this oscillation, as a final step we apply a line-search, for details see supporting information, algorithm~\ref{linesearch}. 
The described gradient descent algorithm provides an efficient way for minimizing Eq~(\ref{map:gamma}).
When calculating the gradient, we neglect the diagonal. Consequently, we also neglect the diagonal of the resulting estimated matrix, so we are not able to study self connections of the nodes.

\begin{table}[!ht]
\centering
\caption{
{\bf Parameter used for estimation}}
\begin{tabularx}{\textwidth}{L{14.5cm} R{1.5cm} }
\multicolumn{1}{X}{\bf Parameter}& \multicolumn{1}{>{\raggedleft}X}{\bf Value}\\ 
\rowcolor{Gray}
xtol & $0.7 \cdot 10^{-2}$ \\ 
ftol & $0.7 \cdot 10^{-4}$ \\ 
\rowcolor{Gray}
gtol & $0.7 \cdot 10^{-2}$ \\ 
$\kappa$& $500$ \\ 
\rowcolor{Gray}
$\lambda$ & $5$ \\ 
\end{tabularx}
\label{table_parameter_est}
\end{table}

\subsubsection*{Initial condition}
As starting condition for the gradient descent we use a matrix $ B_0(f)$ such that
\[
B_0^*(f)  B_0(f) = \hat{C}^{-1}(f).
\] 
There are many ways to choose a $B_0$ with this property, we found the following choice efficient: As $\hat{C}$ is the cross-spectral density it is positive definite, and so is $\hat{C}^{-1}$. Thus, there is exactly one positive definite square root of $\hat{C}^{-1}$ \cite{Horn1985} which can be calculated by 
\begin{equation}\label{eq:B0}
B_0 = \sqrt{\hat{C}^{-1}} = W \sqrt{E} W^* 
\end{equation} 
where the columns of $W$ are the eigenvectors of $\hat{C}^{-1}$, and $E$ is the matrix with the corresponding eigenvalues of $\hat{C}^{-1}$ on the diagonal. Thus we initialize the gradient descent with $U_0 = \mathbbm{1}$ and $B_0$ given by Eq~(\ref{eq:B0}).

\subsubsection*{Step  size selection} 
A critical part of the optimization is the selection of an appropriate step size.
If the step size is too large, one might miss the minimum. If the step size is too small, the optimization converges very slowly.
For the gradient descent, we use an adaptive scheme inspired by \cite{Abrudan2009}, where the step-size depends on the largest eigenvalue of the actual gradient: Let $\lambda_{\max}$ be the largest eigenvalue of $d_\mathrm{act}$, the step size is given by
\[
\delta_\mathrm{act} = \frac{2 \pi}{|\lambda_{\max} | \cdot \kappa}
\]
where $\kappa$ is constant.
The intuition behind that, is that smooth cost functions along a geodesic on the unitary manifold are almost periodic. So the step size should be a fraction of the period of this function. This is achieved by the scaling with the largest eigenvalue, which allows us to take a scale-invariant fraction of this period.

\subsection*{Validation methods}
\subsubsection*{Noise-free covariance matrices}
We assume that the interactions among neuronal populations can be described by a linear model, see Eq~(\ref{eq:cinv}) with $Z = \mathbbm{1}$. 
This model allows us to derive a relation between the connectivity matrix of the network $G$ and the inverse cross-spectral density matrix $\hat{C}^{-1}$ of the measured activity
\begin{equation}\label{eq:cinv_simplified}
 \hat{C}^{-1} = \left( \mathbbm{1} -\hat{G}^* \right) \left( \mathbbm{1} - \hat{G} \right) = \mathbbm{1} - \hat{G}^* - \hat{G} + \hat{G}^*\hat{G}.
\end{equation} 
Given a sampled connectivity matrix $G$ we can calculate the inverse covariance matrix directly using Eq~(\ref{eq:cinv_simplified}).
For all simulations, half of the connections were negative (inhibitory) connections, the absolute strength was the same for all connections and $20$ repetitions were simulated. 
As connectivity profiles we used random Erd\H{o}s-R\'{e}nyi networks.

\subsubsection*{Ornstein-Uhlenbeck processes}\label{methods:ar}

To validate our inference procedure before applying it to the network inference from measurements of neuronal activity we simulated stationary signals. 
Since there is no gold-standard for simulations of fMRI data\cite{Welvaert2014}, we based our simulations on the Ornstein-Uhlenbeck process \cite{Gardiner2004}, which provides a simple linear model for neural activity. 
\begin{equation}
dx(t) = A x(t) dt +  dW(t)
\end{equation}
where $A$ is a matrix and $W$ a Wiener process. In our applications, we parametrize this matrix as $A = \frac{1}{\tau}(G - \mathbbm{1})$ with real-valued connectivity matrix $G$ and time constant $\tau$.
For this process, it is possible to calculate the stationary covariance matrix $\Sigma$ from the continuous Lyapunov equation
\[
\mathbbm{1} = A \Sigma + \Sigma A^T.
\]
In fact, we simulated the process in discrete time.
In analogy with \cite{Gillespie96} we use  a multivariate version of  the exact update formula
\begin{equation}
x(t + \Delta t) = e^{A \Delta t} x(t) + n(t) ,
\end{equation}
where $n(t) \sim N(0, \Sigma)$ is normally distributed, with $\Sigma$ being the stationary covariance matrix described above.
As a final step, we filter the time series $x(t)$ with the canonical hemodynamic response function (HRF) \cite{Friston1998, Glover1999}.
To match the data obtained in brain scans sampled at a temporal resolution of $\Delta t = 0.1\, \mathrm{s}$, we used random connectivity profiles $G$  with a connection probability $p = 0.1$ (Erd\H{o}s-R\'{e}nyi model), $50\%$ negative entries, and a spectral radius of $\rho = 0.3$. All parameters used are listed, once more, in table \ref{table_parameter_sim}.

\begin{table}[!ht]
\centering
\caption{
{\bf Parameter used for simulations}}

\begin{tabularx}{\textwidth}{L{13.5cm} R{2.5cm} }
\multicolumn{1}{X}{\bf Parameter}& \multicolumn{1}{>{\raggedleft}X}{\bf Value}\\ 

\rowcolor{Gray} repetitions & $20$ \\ 
network type & Erd\H{o}s-R\'{e}nyi \\ 
\rowcolor{Gray}N & $100$ \\ 
p & $0.1$ \\ 
\rowcolor{Gray}T & $350\,000 \,\mathrm{s}$ \\ 
dt & $0.1 \,\mathrm{s}$ \\ 
\rowcolor{Gray}$\tau$ & $0.1 \,\mathrm{s}$ \\ 
$\rho$ & $0.3$ \\
\end{tabularx}

\label{table_parameter_sim}
\end{table}

Before calculating the covariance $C$, the data is standardized such that the mean is $0$ and the variance is $1$ for all components of the time series. 
We add normally distributed observation noise  $u_{\mathrm{obs}}$ with a $\mathcal{N}( 0, \sigma_{\mathrm{obs}} )$ distribution to the simulated signal. After the simulation we calculated the signal-to-noise ratio according to 
\[ 
\mathrm{SNR} = \frac{\sigma_X^2}{\sigma_{\mathrm{obs}}^2} 
\]
where $\sigma_X^2$ denotes the variance of the signal.

\subsection*{Performance measures}
When estimating connectivity from simulations with known underlying network structure (ground truth), one can quantify the performance of the estimation.
For measuring the accuracy of our estimation we employ three different methods.

First, we use the area under the ROC-curve (AUC). The ROC (receiver operating characteristic) curve is obtained as following: For each possible parameter value (in our case the threshold for the existence of a connection), the number of true-positives (TP) and false-positives (FP) is used to calculate the true-positive rate (or recall) $\text{TP}/(\text{TP} + \text{FN}) $ and the false-positive rate $ \text{FP}/(\text{FP} + \text{TN}) $.
The ROC curve is then obtained by plotting the true-positive-rate against the false-positive rate.

\begin{figure*}[!h]
\centering
\includegraphics[scale=1]{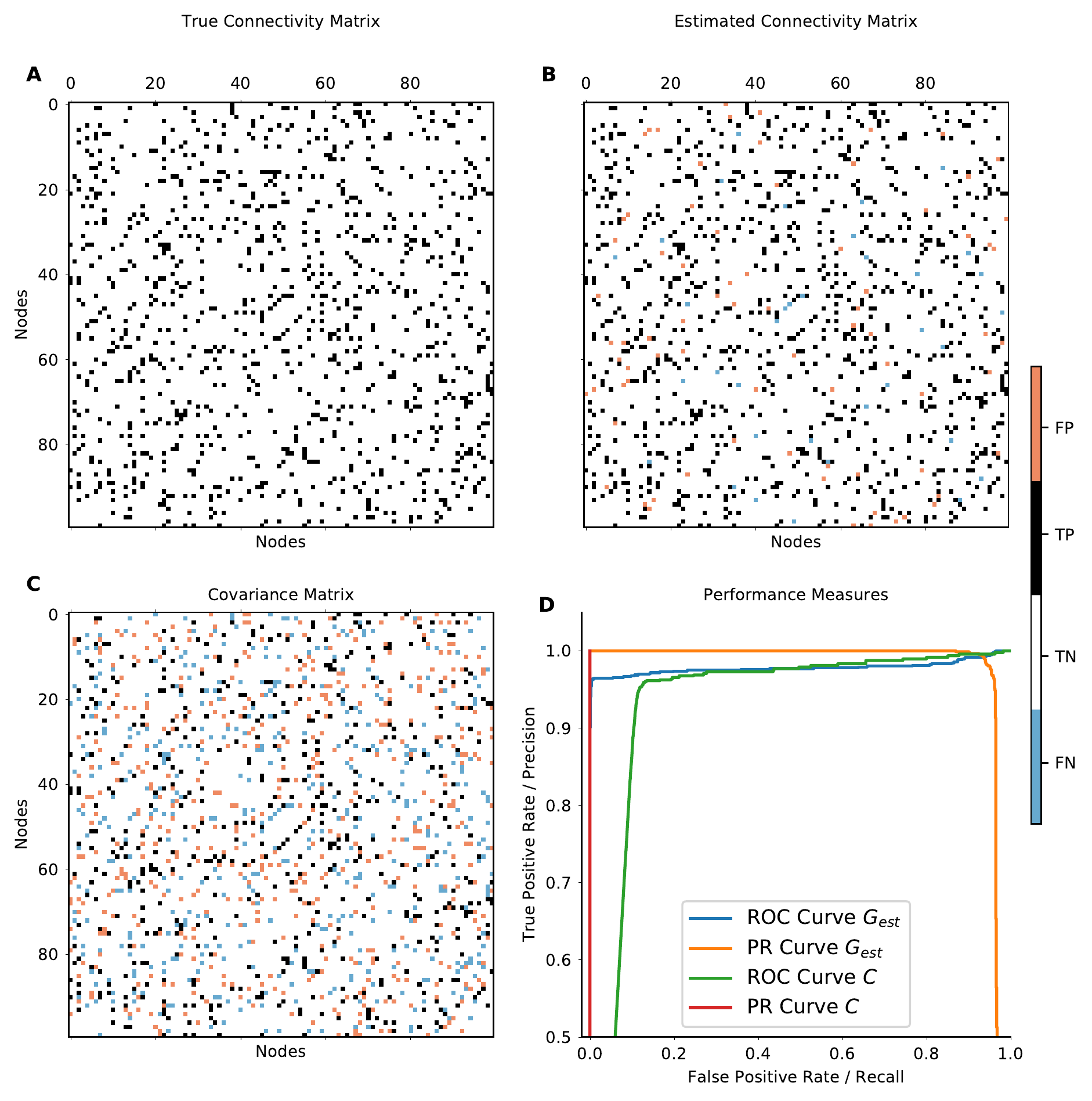}
\caption{Networks inferred from a simulated Ornstein-Uhlenbeck process.$\mathbf{A}$ shows the original network. $\mathbf{B}$ shows the network inferred with our new method from the zero-lag covariances. White and black entries indicate true negative (TN) and true positive (TP) connections, blue and red entries indicate false negative (FN) and false positive (FP) connections, respectively. In this example, the performance measures are $\textrm{AUC} = 0.98$, $\textrm{PRS} = 0.97$ and $\textrm{PCC} = 0.95$. $\mathbf{C}$ depicts the sample covariance (functional connectivity) matrix directly estimated from the data.  In $\mathbf{C}$, as a consequence of symmetry, the number of wrongly estimated connections is quite high, the performance measures are $\mathrm{AUC} = 0.93$, $\mathrm{PRS} = 0.54$, and $\mathrm{PCC} = 0.29$.  $\mathbf{D}$ shows the Receiver Operating Characteristic Curve and the Precision Recall Curve for the networks estimated from zero-lag covariance $G_\mathrm{est}$ in blue/orange and of the functional connectivity $C$ in green/red. The areas under these curves are the AUC and PRS, respectively.}\label{result:ar_GGestGestcsd}
\end{figure*}

Secondly, we use the average precision score (PRS) which is the area under the precision-recall curve. This also includes the false-negatives (FN) (precision: $\frac{\text{TP}}{\text{TP} + \text{FP}}$). 
If both AUC and PRS are equal to $1$, the connections in the network are perfectly estimated. 
Sample curves are shown in Fig~\ref{result:ar_GGestGestcsd} D.

Thirdly, we calculate the Pearson Correlation Coefficient (PCC) which in contrast to the measures defined before also take the strength and the sign of the interactions into account. This also means that this measure is less suited to assess whether a connection exists or not. It rather measures whether the estimated connections have the same strength as the original ones.
We consider all three performance measures simultaneously to establish the quality of our estimates.

\subsection*{Experimental fMRI data}\label{methods:mreg}
Seven healthy subjects underwent a 20-minute resting-state fMRI experiment on a 3 T Siemens Prisma scanner. 
The data was acquired using the MREG sequence \cite{Asslaender2013}, yielding a high temporal resolution ($\mathrm{TR} = 0.1 \, \mathrm{s}, 12000$ time points) that facilitates functional connectivity analyses \cite{LeVan2017}. 
The other sequence parameters were $\mathrm{TE} = 36\,\mathrm{ms}$, $\mathrm{FA} = 25^\circ$, $64 \times 64 \times 50$ matrix and $3\,\mathrm{mm}$ isotropic voxel size.
Additionally, cardiac and respiratory signals were recorded with the ECG and abdominal breathing band  from the scanner's physiological monitoring unit. 
Motion correction was done with FSL and  physiological noise correction  was performed with RETROICOR \cite{Glover2000}. 
Average CSF and white matter signals were regressed out, but no global signal regression was performed. 
Following image normalization to MNI space, voxels were parcellated according to the AAL atlas (excluding the cerebellum), and the mean activity within each atlas region was calculated. 
The connectivity was then estimated using zero-lag covariances of the standardized signals.
\subsubsection*{Ethics statement}
The experiments have been approved by the Ethics Committee of the University Medical Center Freiburg.

\section*{Results}
\subsection*{Noise-free covariance matrices}
Intrinsic properties of our new estimation procedure can be identified by studying the performance of the method for perfectly estimated (noise-free) covariance matrices. 
This way we address properties that do not depend on any particular feature of the underlying data, and that are not due to the success of the measurement process. 
In particular, we show for which types of networks our estimation procedure gives good results on technical grounds, with a wide range of networks hopefully including those arising in applications. We used random Erd\H{o}s-R\'{e}nyi connectivity profiles for all simulations.

The macro-connectivity between neuronal populations has to satisfy certain conditions in order to be tractable by our methods. Two of these conditions concern the dynamic stability of the network and the strength of the interactions. There is a trade-off between the number of physical links and the resulting strength of macro-connections, and the dynamic stability of the network. To study the performance of our method in these various regimes, we separately varied the network size $N$, the connection probability $p$,  and the absolute strength of connections $|J|$ in the connectivity matrix $G$, while the fraction of inhibitory couplings was kept at $50\%$. The spectral radius $\rho$ of the bulk eigenvalue spectrum is approximately given by
\begin{equation}\label{eq:spectralradius} 
\rho^2 = J^2 p (1 - p ) N .
\end{equation} 
The default values of the parameters used in our study were $N = 100$, $p = 0.1$ and $\rho = 0.7$, where only one of them at a time was systematically varied.
Low values of the spectral radius $\rho$ correspond to networks with weak recurrent interaction and high values to networks with strong interaction, respectively. According to the model of network interaction assumed here, the networks need to have a spectral radius $\rho > 0$ for network interaction to be present and $\rho < 1$ for the dynamics to be stable. 

\begin{figure*}[!ht]
\begin{center}
\includegraphics[scale=1]{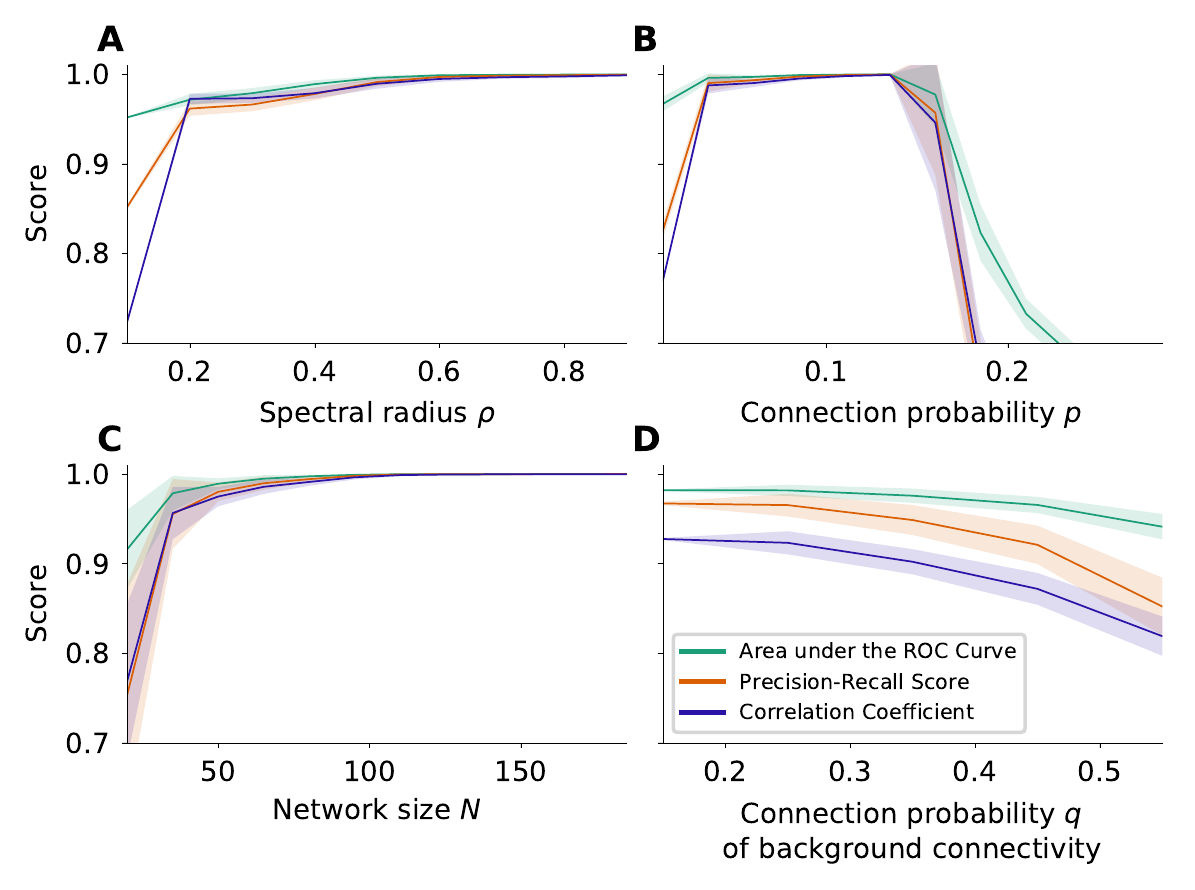}
\caption{Effects of spectral radius $\rho$, connection probability $p$ and network size $N$. Here we consider the case of noise-free covariance matrices, which were created based on the theory of the underlying model. The quantities considered are the area under the ROC curve (AUC; green), the precision recall score (PRS; orange) and the Pearson correlation coefficient (PCC; purple). The shaded areas indicate the mean $\pm$~standard deviation computed over $20$ realizations. $\mathbf{A}$ If the network interaction is larger than $\rho_\mathrm{min}$, it has relatively little effect on the performance of the estimation. Even in the extreme case, where $\rho$ is close to $1$, the estimation works well. $\mathbf{B}$ Performance of the estimation for different sparsity levels, encoded by the respective connection probabilities $p$. As expected, for non-sparse networks the performance of the algorithm degrades dramatically. $\mathbf{C}$ Performance of the estimation for increasing network size. Our results indicate clearly that bigger networks can be better reconstructed. Applicability may be limited by the numerical effort associated with the optimization. $\mathbf{D}$ Performance of the estitmation in presence of weak background connections. It is nevertheless possible to infer the skeleton of strong connections with high fidelity. }\label{result:clean}
\end{center}
\end{figure*}

First, our results in Fig~\ref{result:clean}~A indicate that a certain minimal level of interaction is necessary to be able to estimate the connections reliably. 
Above a value of $\rho_\mathrm{min} = 0.2$, the influence of the spectral radius on the performance of the estimation is weak, but the larger the spectral radius is the better the estimation gets.

Secondly, the connection probability of the network influences the quality of the estimation. 
For all connection probabilities tested here the network size was kept constant at $N = 100$ nodes. The networks were constructed such that the strength $|J|$ of all connections was the same and such that the spectral radius $\rho$ was constant according to Eq~(\ref{eq:spectralradius}). 
Fig~\ref{result:clean} B shows that the estimation works very well for sparse matrices with a connection probability in the range between $5\%$ and $15\%$.
For networks with higher connection probability and equally strong connections, the performance decreases as expected, due to the bias associated with $L^1$-minimization. 
But even for a connection probability of $p = 0.21$, a fraction of $14.2\%$ of the estimated connections are false negative, and $3.3\%$ are false positive.
More than $90\%$ of the correctly estimated connections have the correct sign.
In applications, the focus of the estimation often lies on the strongest connections in the network. 
In networks with a background of weak connections and a sparse skeleton of stronger connections, it is possible to selectively estimate these strong links although, strictly, the assumption of a sparse network is violated.
Fig~\ref{result:clean} shows the performance of our method for such networks: the networks consist of a skeleton of strong connections with connection probability $p = 10 \%$ and a connection strength derived from Eq~(\ref{eq:spectralradius}) for $\rho = 0.7$. 
Additionally, we created a second network with weaker connections for various connection probabilities $q$. 
The two networks were combined by adding the connectivity matrices. 
The connection strength of this weaker connections is also derived from Eq~(\ref{eq:spectralradius}), with a spectral radius of the background network being $20 \%$ of the spectral radius of the skeleton network.
Then the performance of the estimation is calculated with respect to the skeleton of strong connections.

Thirdly, to be applicable to a broad range of data types, a method of connectivity estimation should perform stable for different network sizes $N$.
For most common types of non-invasive recordings of population activity the number of nodes considered is in the range between $30$ and $150$.  
It is, of course, possible to consider larger networks, although the estimation becomes computationally more expensive. The runtime of the algorithm for networks with $200$ nodes still in the range of seconds on a state-of-the-art desktop computer, but even networks with $1\,000$ nodes or more are tractable.
The strength of the connections $|J|$ are set such that the spectral radius $\rho$ of $G$ is constant; the connection probability is constant at $p = 0.1$. 
Fig~\ref{result:clean}~C shows that our method performs better for bigger networks. 
We have observed that the $L^1$ cost landscape becomes smoother for larger networks.

\subsection*{Ornstein-Uhlenbeck processes as model for BOLD signals}\label{section:ou}

\begin{figure*}[!h]
\begin{center}
\includegraphics[scale=1]{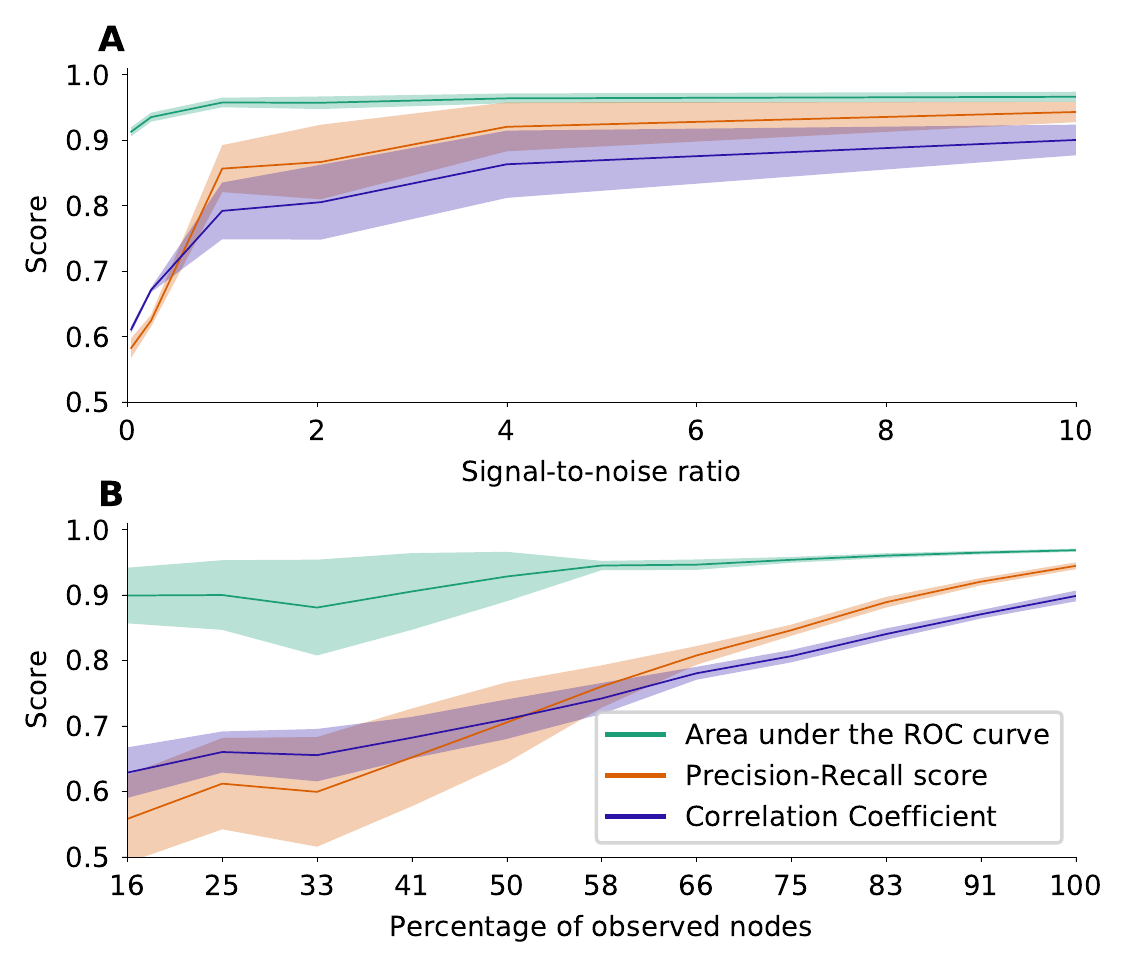}
\caption{Performance of network inference based on simulated Ornstein-Uhlenbeck processes. Same colors as in Fig~\ref{result:clean}. $\mathbf{A}$ Performance of the estimation when measurement noise is added. $\mathbf{B}$ Performance of the estimation if only parts of the network are observable. The fraction of observed nodes in a network are indicated on the $x$-axis. The total number of nodes in the network was $N = 180$.
}\label{result:ar}
\end{center}
\end{figure*}

In order to create surrogate data which fit fast fMRI data \cite{Asslaender2013}, we simulated interacting stochastic processes known as Ornstein-Uhlenbeck processes. 
In this case, the performance of the network inference depends on how well the inverse covariance matrix, which is the basis of the estimation, can be derived from the data. 
In addition to finite size effects, we studied the impact of observation noise on the performance, see Fig~\ref{result:ar}.
We used $N = 100$, $p = 0.1$, $dt = 0.1 \,\mathrm{s}$, $\rho = 0.74$  and $\tau = 0.1 \,\mathrm{s}$ as default values of the parameters.
Generally, it seems natural to use Welch's method to calculate cross-spectral densities directly, and then to estimate the connectivity for each frequency band separately.
For the data described here, however, we can estimate the connectivity from zero-lag sample covariances in the time domain.
This is possible when the mass of the covariance function is concentrated very close around lag $0$. 
Then lag $0$ is the only one contributing to the integral of the covariance function, which corresponds to the cross-spectral density $\hat{C}(0)$.

As shown in Fig~\ref{result:ar} A, with noisy data the AUC is still good, but the PRS is lower than in the case, where the covariance is known without error. However, for a signal-to-noise ratio above $1$  the performance improves very quickly.

In the case of fMRI usually the whole brain is scanned, and there are no unobserved nodes in the network. 
However, for other data types (e.g.\ fNIRS) only parts of the brain can be observed. 
The question then is, whether this sub-network can nevertheless be reconstructed from the recorded signals. 
To model this scenario, we took simulated data and removed randomly a certain subset of components from the dataset.
The interaction of the removed nodes is then not part of the covariance matrix of the reduced dataset, although the unobserved nodes of course still exert their influence on the observed ones.
The performance of the estimation of the sub-network based on the reduced dataset is shown in Fig~\ref{result:ar} B. 
Our analysis shows very clearly that the estimation still leads to reasonable results under these conditions.
In fact, we can demonstrate that we are inferring causal connections only: For unconnected observed nodes $X$, $Y$ and a latent node $L$ connected to both $X$ and $Y$, our method does not erroneously indicate a link between $X$ and $Y$.

\begin{figure*}[!ht]
\begin{center}
\includegraphics[scale=1]{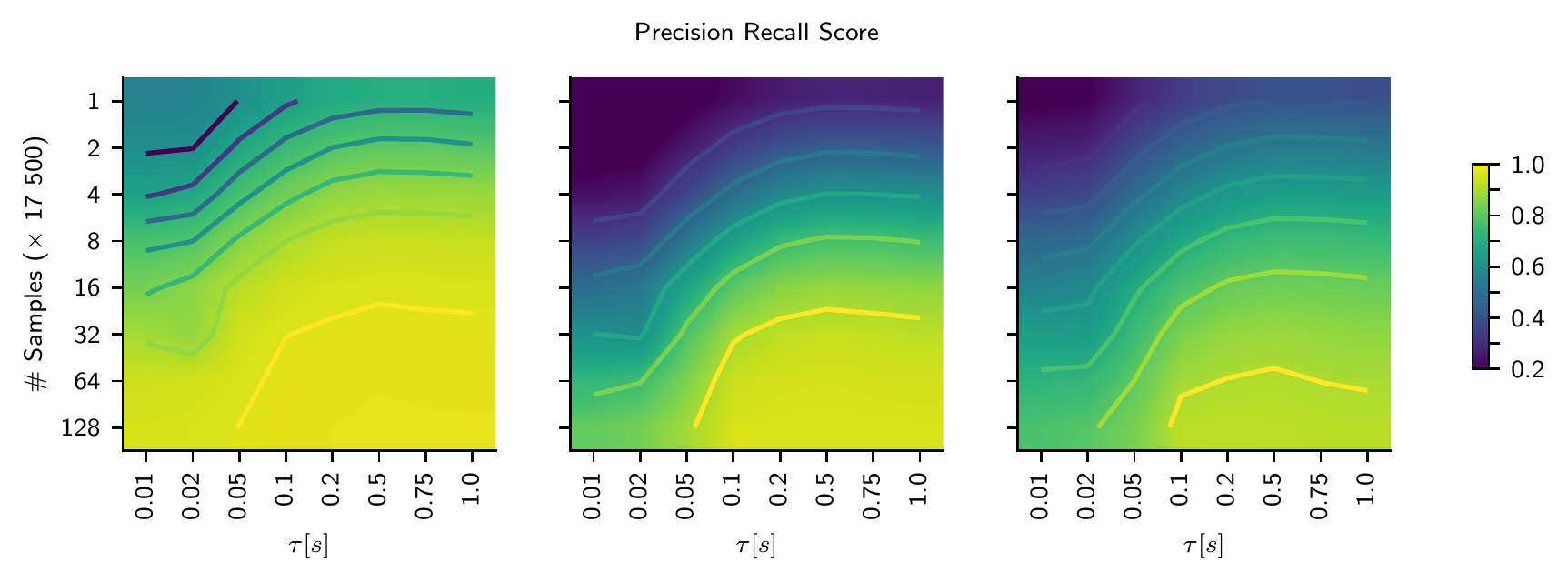}
\caption{Performance (color coded) of the estimation depending on data length ($y$-axis) and time scale of the activity ($x$-axis). Both scales are logarithmic. For interpolation a bilinear method is used.
}\label{result:laenge_tau}
\end{center}
\end{figure*}

One key factor for a reliable estimation of the covariance matrix is the amount of data available. 
This depends on the length of the measurement or simulation, and on the sampling rate. 
Since fast fMRI time series are obtained by measuring the BOLD response as a proxy of neuronal activity, the time scale of the measured data is relatively slow compared to the time scale of the underlying neuronal activity. 
Fig~\ref{result:laenge_tau} shows the performance of network inference depending on the amount of data available, and on the time-scale of the neuronal activity.
Not surprisingly, the more voluminous the dataset is, the better the estimation gets.
On the other hand, it shows that the estimation generally leads to better results for slower temporal dynamics.
Also, for data of sufficient length with a fairly good signal-to-noise ratio, the estimation of the connectivity is possible even when only a part of the network is observed.
To allow comparison of our new method with other known methods for network inference \cite{Nie2017, Ryali2016, Hyvarinen2013}, we applied it to the \textit{NetSim} dataset provided by \cite{Smith2011}.
For details on the result of this, please see Fig~\ref{result:netsim} in the supporting information. 

\subsection*{fMRI data}
\begin{figure*}[!ht]
\begin{center}
\includegraphics[scale=1]{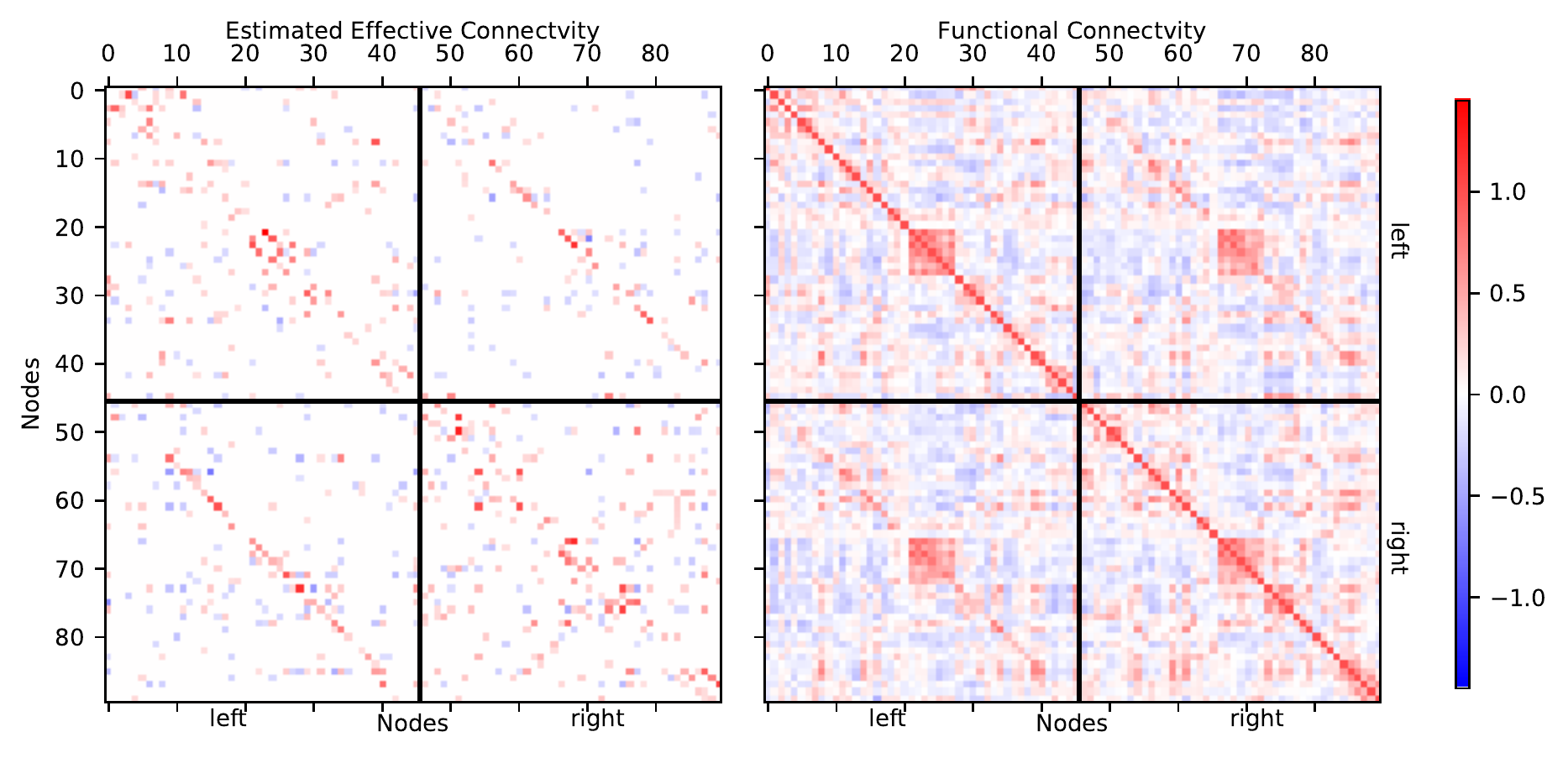}
\caption{Left panel: Directed connectivity estimated with our new method from one sample MREG data set. Voxels were parceled using the AAL90-atlas. In the top-left block of the connectivity matrix connections within the left hemisphere are shown, in the lower-right block connections within the right hemisphere. The off-diagonal blocks represent the inter-hemispheric connections from the left to the right hemisphere (lower left) and from the right to the left hemisphere (top right). The strength of all connections is color coded, with red representing positive (excitatory) connections and blue representing negative (inhibitory) connections. Only the strongest $10\%$ of connections are shown. Right panel: Functional connectivity matrix derived from the same data. 
}\label{result:MREG}
\end{center}
\end{figure*}

We estimated connectivity from seven fast fMRI datasets, for details see the methods section.
The resulting networks, after a threshold of $10\%$ was applied, consist of $810$ connections for each dataset. 
The threshold of $10\%$ was chosen arbitrarily. In the supporting information (Fig~\ref{result:MREG_hist}) we show the histogram of estimated connection strengths for all seven reconstructed networks before thresholding. The threshold is derived from the $10\%$ strongest connections, disregarding their signs.
As there is generally no full ground truth for the connectivity inferred from human fMRI recordings available \cite{Zaghlool2014, Nie2017, Ryali2016}, we cannot definitely assess the degree to which the result of our inference are correct. We can, however, establish whether they are plausible.
One representative connectivity matrix is shown in Fig~\ref{result:MREG}. On average, $34\%$ of the connections were inhibitory, with  negligible variability across subjects.
Of all connections found, $301 (37\%)$ were found in four subjects or more, and $4\,872$ out of $8\,100$ possible connections were absent in all subjects.
On average, $245$ of the connections were bi-directional and $565$ connections were identified only for one direction. 
In general, close-by areas are more likely to be connected than more distant ones. This fact is (approximately) represented by a concentration of connections along secondary diagonals in the within-hemisphere blocks. 
Also, there are frequent inter-hemispheric connections between corresponding areas. This fact is represented by the diagonal entries in the across-hemisphere blocks.
\subsection*{Comparison with the Regularized Inverse Covariance (RIC) method}
\begin{figure*}[!ht]
\begin{center}
\includegraphics[scale=1]{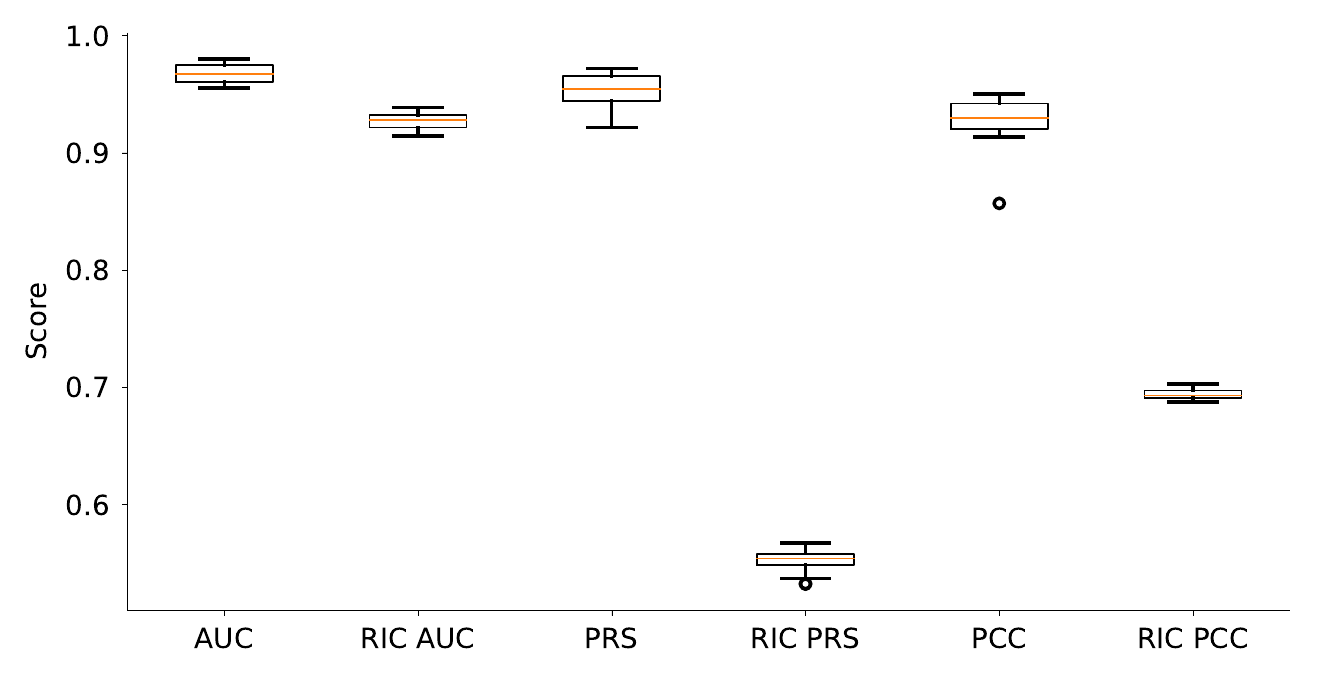}
\caption{Comparison of performance with the Regularized Inverse Covariance (RIC) method based on numerical simulations of Ornstein-Uhlenbeck processes. Shown are the results from the reconstruction of $20$ different networks with Erd\H{o}s-R\'{e}nyi connectivity profiles as described before (cf. Fig~\ref{result:ar_GGestGestcsd}). AUC, PRS and PCC of our new method and of the RIC method, respectively, are shown side-by-side.
}\label{result:comparison}
\end{center}
\end{figure*}

As mentioned above, different heuristics have been suggested to reconstruct networks from neuronal signals. 
In Fig~\ref{result:comparison} we compare the performance of the new method we propose here and the established method of Regularized Inverse Covariance \cite{Smith2011}, based on the implementation provided at \texttt{https://fsl.fmrib.ox.ac.uk/fsl/fslwiki/FSLNets}.
Our comparison clearly shows that our new method performs significantly better than the Regularized Inverse Covariance method, mainly, because the latter cannot establish the direction of connections.
The superior performance of the new method is reflected in higher values for all three performance measures, in particular PRS and PCC.
As regularization parameter required by the software toolbox, we used $\lambda = 5$. 

\begin{figure*}[!ht]
\begin{center}
\includegraphics[scale=1]{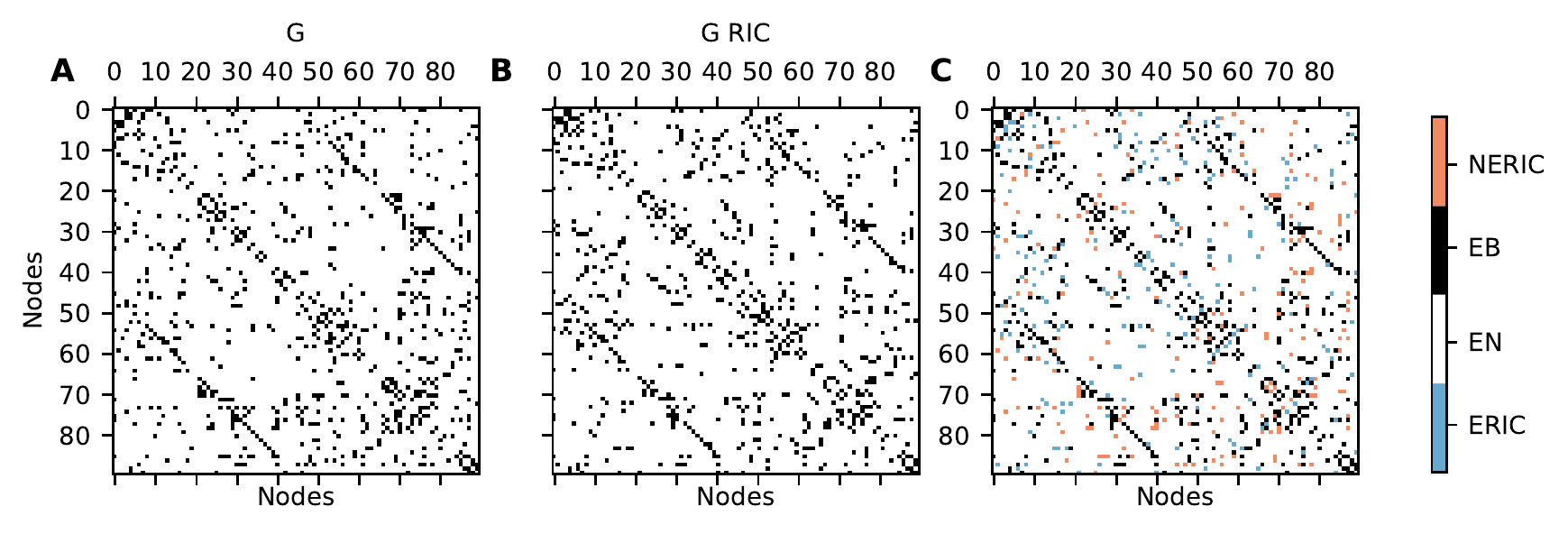}
\caption{Estimated networks for one representative MREG dataset. The left panel shows the symmetrized network reconstructed with our estimation method, the middle panel shows the network found with the RIC method. The right panel shows the connections which are identified by both methods (EB, black), by none of the methods (EN, white), the connections found only by the RIC (ERIC, blue) and the connections found only by our method, but not by the RIC method (NERIC, red).
}\label{result:comparison_MREG}
\end{center}
\end{figure*}

Furthermore, we applied the RIC method on all seven MREG datasets described before.
A threshold was applied, such that only the 10$\%$ strongest connections are retained.
To compare the outcome of both methods, we only condidered the existence of connections (binary and symmetric connectivity) and disregarded weights and directions (weighted nonsymmetric connectivity).
One representative example of the comparison of both methods is shown in Fig~\ref{result:comparison_MREG}.
For RIC, $376$ out of $810$ possible connections where identified in four subjects or more out of seven, the corresponding number for our method is $392$ out of $810$ possible connections. If any method produced directed networks with $10\%$ connection probability at random, this would yield an average count of less than 25 connections ($3 \%$ of $810$ connections) that agree for least four out of seven independently generated networks.
On average, $290.5$ out of $4\,050$ possible connections (undirected) are identified by both methods, $3\,530.5$ connections were found by neither of the methods. This means that both methods agree on $3\,821$ out of $4\,050$ connections on average. The two methods disagreed on the remaining $229$ connections.

\section*{Discussion}

With the described method we can estimate directed and signed effective connectivity between neural populations from measured brain signals, based on zero-lag covariances only. 
To investigate the reliability of our estimated connections we used simulations of Ornstein-Uhlenbeck processes mimicking BOLD-related signals generated by interacting neuronal populations. 
Our method shows very good performance, if enough data is available and the observation noise is not too strong. 
Also, even in cases with relatively poor performance (e.g.\ if the network is too dense) more than $90\%$ of the estimated connections have the correct sign. 
Applying the method on measured fast fMRI data, we found that about $34\%$ of all identified connections have an inhibitory effect on their respective target population.
In general, inhibitory synapses are mainly formed within local populations, and typically do not project to distant targets. 
An inhibitory connection between populations, however, can also be achieved by excitatory neurons preferentially terminating on the inhibitory neurons of the target region.
The comparison with the Regularized Inverse Covariance (RIC) method shows good agreement with regard to the existence of connections. Directions cannot be disambiguated with the RIC method.
Our results based on simulated surrogate data reflect what one would expect from the design of an estimation procedure. 
For large, sparse networks with sufficiently strong interaction, our network estimation procedure works reliably. 
However, as expected if the network is not sparse, or the time series is too short, the quality of the estimate drops.
Nevertheless, in most cases the main interest lies on the strongest connections, which can be reliably estimated with our method even when the network is not sparse. 
For the experimental data shown, individual connections may be unreliable because of the limited size of the dataset.
Also, it is unclear whether the biological network to be analyzed is really sparse, and if the assumption of pairwise independent external input is really justified.
On the other hand, due to the higher likelihood of a coupling between close-by areas and between inter-hemispheric counterparts, the resulting network looks plausible.
For interpreting individual connections longer recordings would certainly be beneficial. 
Also, one could then use temporal information from additional frequency bands. 
Of high interest is also the comparison with structural measures as the ones obtained by diffusion tensor imaging.
To the best knowledge of the authors, this is the first time that effective whole-brain connectivity has been estimated from zero-lag covariances. 
Other methods \cite{Gilson2016} rely on lagged covariances, where the correct lag parameter is critical, and needs to be inferred from the exponential decay of the observed auto-covariances.
Also, our proposed method is the only one that can detect directed inhibitory connections on the whole-brain scale. 
The estimation procedure is fast and easy to apply.
As it uses no temporal information, our method can also be applied on other data types that rely on the BOLD effect, e.g.\ fNIRS, but also data types measuring electrical population activity directly.
This makes it a good candidate for, among other things, studying changing connectivity in neurodegenerative diseases, like Parkinson's or Alzheimer's.

\section*{Conclusion}
With the presented method we can estimate directed effective connectivity on a whole-brain scale.
Also we are able to detect whether connections are excitatory or inhibitory.
The estimation is possible based on zero-lag covariances, but can also be applied to frequency-resolved cross spectral densities.

\section*{Supporting information}
\quad
\begin{algorithm}[!ht]
\caption{Conjugate gradient descent}\label{algo:conj}
\begin{algorithmic}[1]

\State initialize $U, B_0$
\State calculate gradient $d$ and $a \longleftarrow \frac{1}{2}(d - d^\ast)$
\State calculate step size $\delta$
\State $U \longleftarrow \exp(-\delta a)$
\While {not converged}
\State $a^\prime \longleftarrow a$
\State calculate gradient $a$
\State $\beta \longleftarrow \frac{\langle a, a + a^\prime \rangle}{\langle a^\prime, a^\prime \rangle}$
\State $ g \longleftarrow -a - \beta a^\prime $
\State calculate step size $\delta$ of $g$
\State $U \longleftarrow \exp(\delta g)$
\EndWhile 
\State Linesearch($U, B_0$)
\end{algorithmic}
\end{algorithm}

\begin{algorithm}[!ht]
\caption{Line Search with Armijo step size rule}\label{linesearch}
\begin{algorithmic}[1]

\Function{LINESEARCH}{$U_\mathrm{act}, d, \alpha$}
\Comment current estimate $U_\mathrm{act}$, gradient $d$, initial-step length $\alpha$
\State $U \longleftarrow exp(-\alpha d)$
\State $Q \longleftarrow UU$
\While {$\Gamma(U_\mathrm{act}B_0) - \Gamma(QB_0) \geqslant \alpha \langle d, d \rangle$}
\State $U \longleftarrow Q$
\State $Q \longleftarrow UU$
\State $\alpha \longleftarrow 2\alpha$
\EndWhile
\While {$\Gamma(U_\mathrm{act}B_0) - \Gamma(UB_0) \geqslant 0.5 \alpha \langle d, d \rangle$}
\State $U \longleftarrow exp(-\alpha d)$
\State $\alpha \longleftarrow 0.5\alpha$
\EndWhile
\Return $U, \alpha$ 
\EndFunction
\end{algorithmic}
\end{algorithm}

\includegraphics[scale=.85]{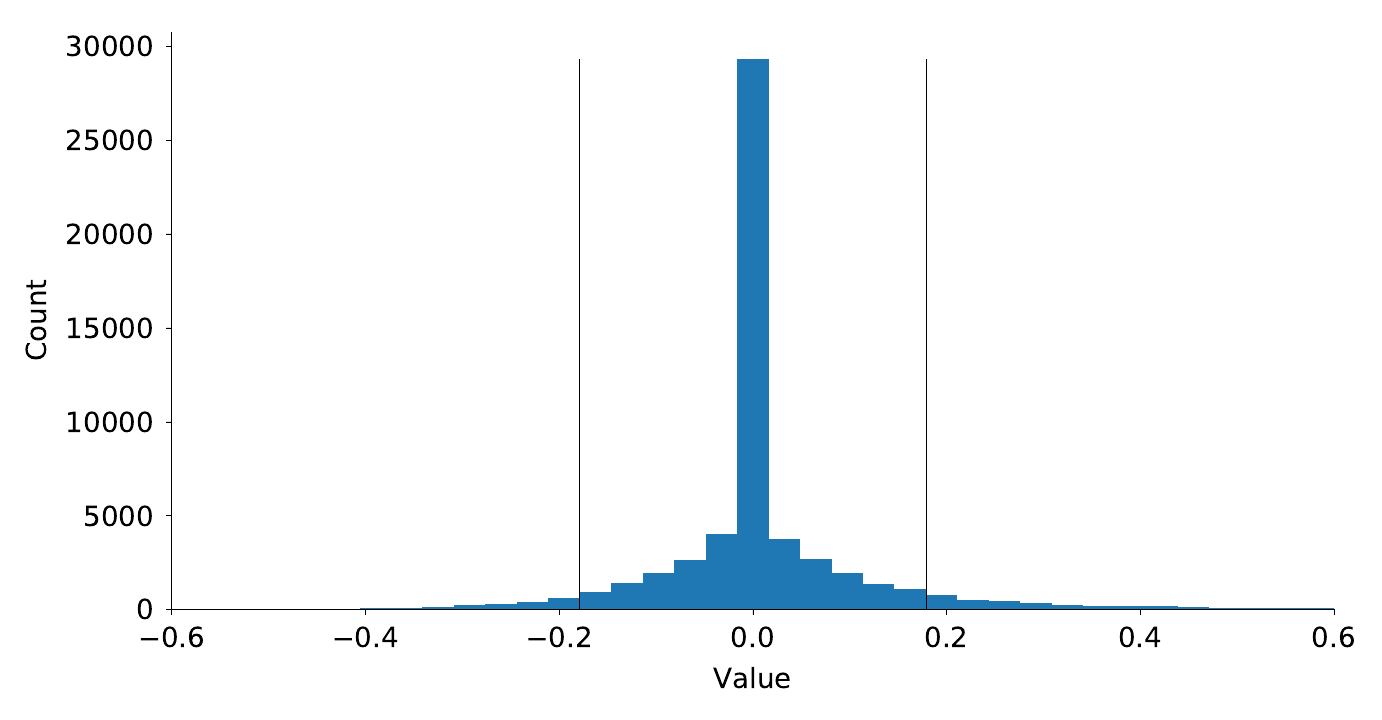}
\captionof{figure}{Histogram of estimated connection strengths taken from the reconstructed networks of all seven subjects. The vertical lines show the thresholds for the excitatory and inhibitory connections, respectively. Only a part of the histogram is shown, the actual range of values is between $-0.99$ and $2.54$.
}\label{result:MREG_hist}
\includegraphics[scale=.85]{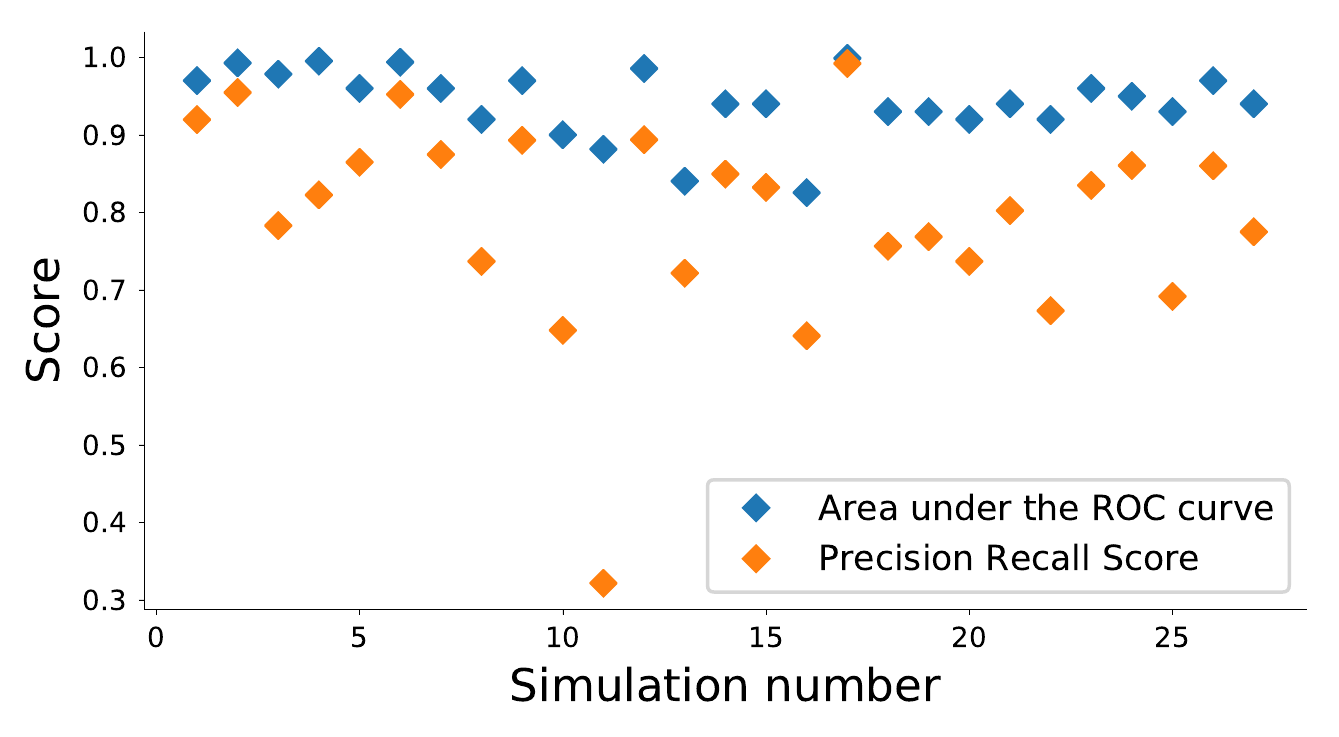}
\captionof{figure}{Performance of our new inference method on the \textit{NetSim} dataset published by \cite{Smith2011}.
Other methods have also been tested on these simulated data sets \cite{Nie2017, Ryali2016, Hyvarinen2013}.
The $x$-axis represent the indices of simulated data sets, as in the original publication. The $y$-axis shows the AUC and PRS of our estimations.
We estimated the connectivity for every individual subject and applied a threshold of $50 \%$, the resulting networks were then averaged over all available subjects/trials. 
Although the networks considered in this paper cover a range of parameters, where we found that our method performs sub-optimally (the networks are generally too small), it still performs reasonably well on these synthetic data.
We obtained average values for AUC and PRS of $0.94$ and $0.79$, respectively.
}\label{result:netsim}

\section*{Acknowledgments}

Supported by the DFG (grant EXC 1086). The HPC facilities are funded by the state of Baden-Württemberg through bwHPC and DFG grant INST 39/963-1 FUGG. We thank Uwe Grauer from the Bernstein Center Freiburg as well as Bernd Wiebelt and Michael Janczyk from the Freiburg University Computing Center for their assistance with HPC issues.

\quad \\

\end{document}